\documentclass[conference]{IEEEtran}
\usepackage[pdftex]{graphicx}
\graphicspath{{../pdf/}{../jpeg/}}
\DeclareGraphicsExtensions{.pdf,.jpeg,.png}

\usepackage[cmex10]{amsmath}
\usepackage{amsfonts} 
\usepackage{amssymb}
\usepackage{algorithmic}
\usepackage{array}
\usepackage{mdwmath}
\usepackage{mdwtab}
\usepackage{eqparbox}
\usepackage{url}
\usepackage{float}
\DeclareUnicodeCharacter{2212}{-} 

\makeatletter
\g@addto@macro\@floatboxreset\centering
\makeatother

\begin{document}

\title{\LARGE \textbf{Prediction of Space Weather Events through Analysis of Active Region Magnetograms using Convolutional Neural Network}}

    \author{\authorblockN{Shlesh Sakpal}
    \authorblockA{Freedom High School}}

\maketitle

\begin{abstract}
Although space weather events may not directly affect human life, they have the potential to inflict significant harm upon our communities. Harmful space weather events can trigger atmospheric changes that result in physical and economic damages on a global scale. In 1989, Earth experienced the effects of a powerful geomagnetic storm that caused satellites to malfunction, while triggering power blackouts in Canada, along with electricity disturbances in the United States and Europe. With the solar cycle peak rapidly approaching, there is an ever-increasing need to prepare and prevent the damages that can occur, especially to modern-day technology, calling for the need of a comprehensive prediction system. This study aims to leverage machine learning techniques to predict instances of space weather (solar flares, coronal mass ejections, geomagnetic storms), based on active region magnetograms of the Sun. This was done through the use of the NASA DONKI service to determine when these solar events occur, then using data from the NASA Solar Dynamics Observatory to compile a dataset that includes magnetograms of active regions of the Sun 24 hours before the events. By inputting the magnetograms into a convolutional neural network (CNN) trained from this dataset, it can serve to predict whether a space weather event will occur, and what type of event it will be. The model was designed using a custom architecture CNN, and returned an accuracy of 90.27\%, a precision of 85.83\%, a recall of 91.78\%, and an average F1 score of 92.14\% across each class (Solar flare [Flare], geomagnetic storm [GMS], coronal mass ejection [CME]). Our results show that using magnetogram data as an input for a CNN is a viable method to space weather prediction. Future work can involve prediction of the magnitude of solar events.
\end{abstract}
\hfill\break

\IEEEoverridecommandlockouts

\IEEEpeerreviewmaketitle

\section{Introduction}

\subsection{Background and Context}
Earth’s atmospheric conditions are heavily influenced by its surroundings. The conditions in the environment surrounding Earth, referred to as space weather, serve a crucial role in determining its composition and balance. Phenomena such as solar flares, coronal mass ejections (CMEs), and geomagnetic storms have the potential to inflict great harm on infrastructure and communication systems on Earth. 

In 1989, Earth was struck by a devastating geomagnetic storm, resulting in electricity grids being disrupted in the United States and Europe, while cities in Canada experienced complete blackouts \cite{storm}. To mitigate the impact of such an event occurring in modern day, it is necessary to implement a system which forecasts space weather and its effects on the atmosphere. This will allow humans to have greater preparation for the effects of hazardous space weather events.

\begin{figure}
    \centering
    \includegraphics[width=\linewidth]{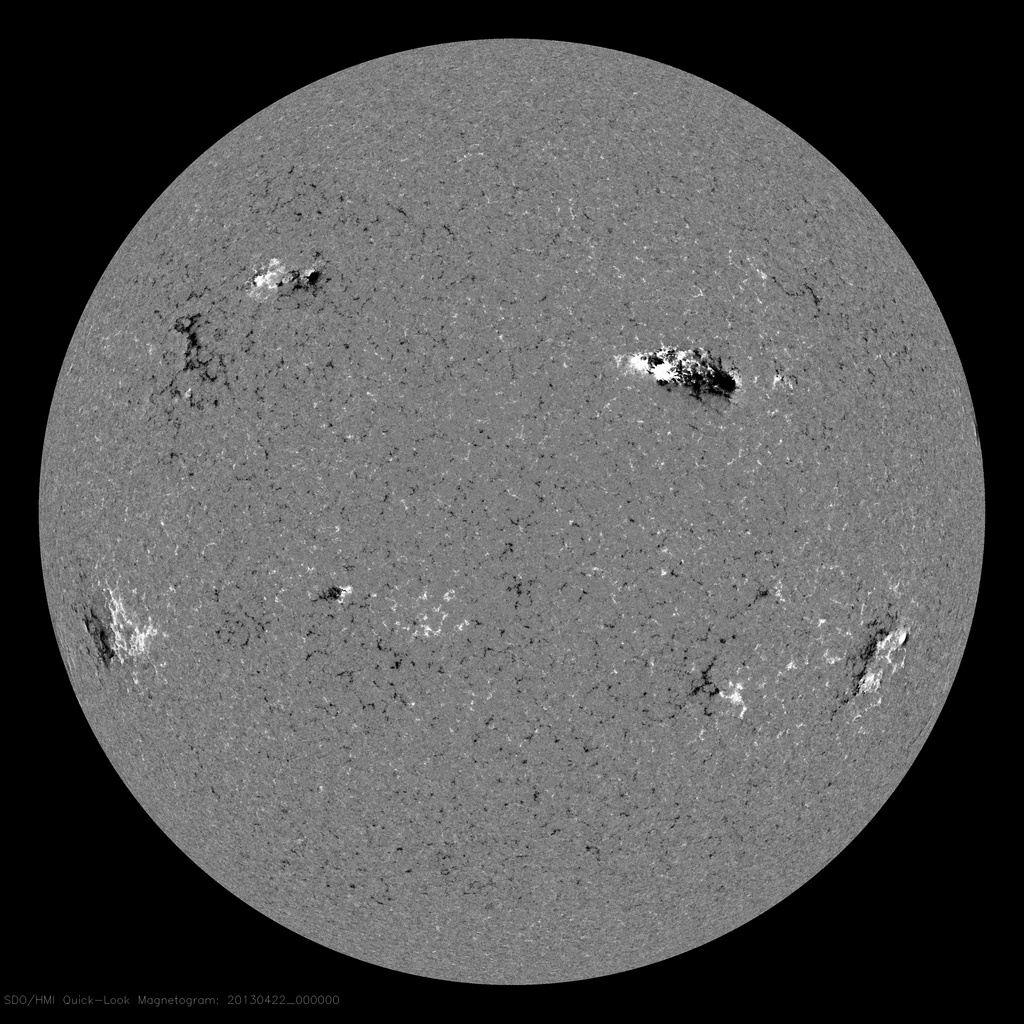}
    \caption{Magnetogram of the Sun. Courtesy of NASA/SDO and the AIA, EVE and HMI science teams.}
    \label{fig:mag}
    \vspace{-\baselineskip}
\end{figure}
The focus of this project will be to develop a system which can predict instances of space weather, with specific focus on solar flares, geomagnetic storms, and coronal mass ejections. The sun experiences a phenomenon called differential rotation, where the Sun’s poles rotate at varying rates when compared to its equatorial regions \cite{rotation}. Due to this property, the magnetic field lines on the Sun become “tangled”, producing regions of greater magnetic activity, known as ‘active regions’. These active regions are where solar flares and coronal mass ejections originate, as shown in fig. \ref{fig:mag}, where coronal mass ejections can cause charged particles from the Sun to interact with Earth’s magnetosphere, creating geomagnetic storms \cite{gms}. 

Because solar events are caused by an increase in charged particles, magnetogram analysis of the Sun has become a more prevalent method to predict instances of space weather. Magnetograms portray the active regions of charged particles on the Sun, where solar events occur \cite{magnetograms}.


\section{Current State of Space Weather Prediction}

\subsection{Geomagnetic Storm}

Geomagnetic storms occur when the charged particles emitted from the Sun experience interactions with Earth’s magnetosphere \cite{gmsforecast}. Current work in predicting geomagnetic storms involves reliance on measurements from ground-based magnetometers and real-time measurement of solar wind \cite{gmsforecast}. A ground-based magnetometer measures alterations of Earth’s magnetic field, allowing research conducted using these instruments to provide insight on when geomagnetic storms occur \cite{gmsforecast}. 

Domico et al. utilized computer vision techniques, including canny edge detection \cite{cannyedge} and topological structure analysis \cite{topological} in order to identify sunspots through images of the Sun, obtained from the NASA Solar Dynamics Observatory \cite{SDO} \cite{gmsforecast}. Analysis of past and present sunspot images, allowed for accurate prediction of geomagnetic storms through a Gaussian Kernel Support Vector Machine, or G-SVM \cite{GSVM}.

\subsection{Solar Flare}

Solar flares are typically detected through active regions (ARs) on the surface of the sun \cite{flareprediction2019}. Recent studies have involved the use of deep learning \cite{deeplearning} in solar flare prediction, a switch from previous statistical analysis. Current work has utilized data collected from the Helioseismic and Magnetic Imager (HMI) of the NASA Solar Dynamics Observatory \cite{flareprediction2022}. The HMI instrument provides data in the form of magnetograms, capturing how the magnetic fields on the Sun’s surface transform over time \cite{HMI}.  The work involving the HMI instrument utilized a convolutional neural network based on the VGGNet model, which allows for the extraction of features in magnetograms to predict between a flare and no-flare case, as well as its associated flare class (C, M, X) \cite{Flareclass}.

\subsection{Coronal Mass Ejection}

Coronal mass ejections (CMEs), generally associated with solar flares, occur when the magnetic field of space realigns explosively, sending a cloud of particles into space \cite{CMEprediction}. Current work explores the application of machine learning into the prediction of CMEs\cite{CMEprediction}. This work focused on determining the relationship between CMEs and Solar flares. The study leveraged data from the HMI instrument, using images of the Sun since 2010, specifically focused on the photospheric magnetic field in order to generate predictions. After classifying the images as CMEs or non-CMEs, the data was then inputted into a Support Vector Machine, which is a type of artificial intelligence model that can identify nonlinear relationships between variables \cite{SVM}.  

\section{Methods and Materials}

\begin{figure}
    \centering
    \includegraphics[width=\linewidth]{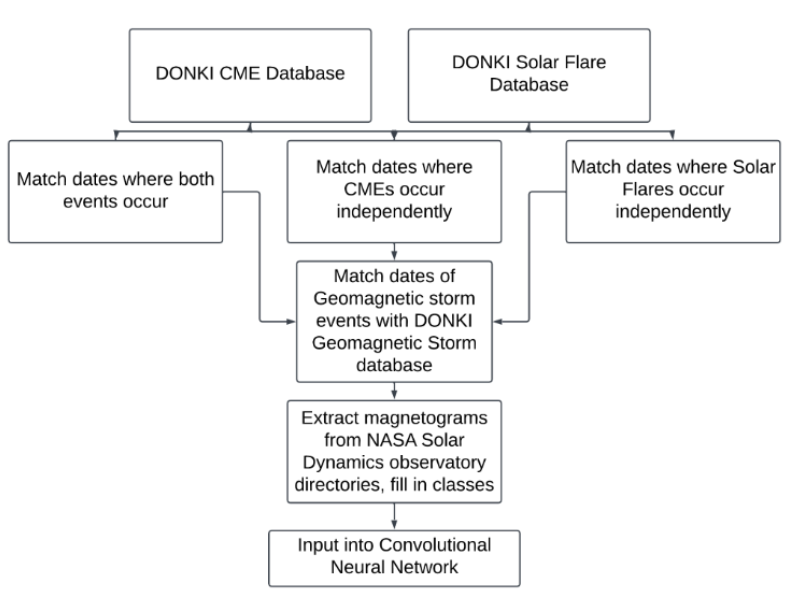}
    \caption{Visualization of overall methodology. Dates between solar flare and CME events were first matched using DONKI database. Then, labels and images were automatically uploaded into dataframe using Google AppsScript which obtained dates from SDO.}
    \label{fig:methods}
    \vspace{-\baselineskip}
\end{figure}

\subsection{Sources}
\vspace{-\baselineskip}
We use the NASA Solar Dynamics Observatory (SDO) and the NASA Space Weather Database Of Notifications, Knowledge, Information (DONKI) \cite{DONKI} as main sources of data for magnetograms and dates. Project was completed entirely digitally, through the use of a personal computer.

\subsection{Data Compilation}
We aggregate solar magnetograms from the SDO, and geomagnetic storm (GMS), solar flare, and coronal mass ejection (CME) arrival times from DONKI (Wold 2018). SDO provides magnetograms at specific dates, which provide visualization of the Sun’s active regions on each day \cite{magnetograms}, and DONKI provides specific timings of all solar events recorded (Wold 2018). 

\begin{figure}[H]
    \centering
    \includegraphics[width=\linewidth]{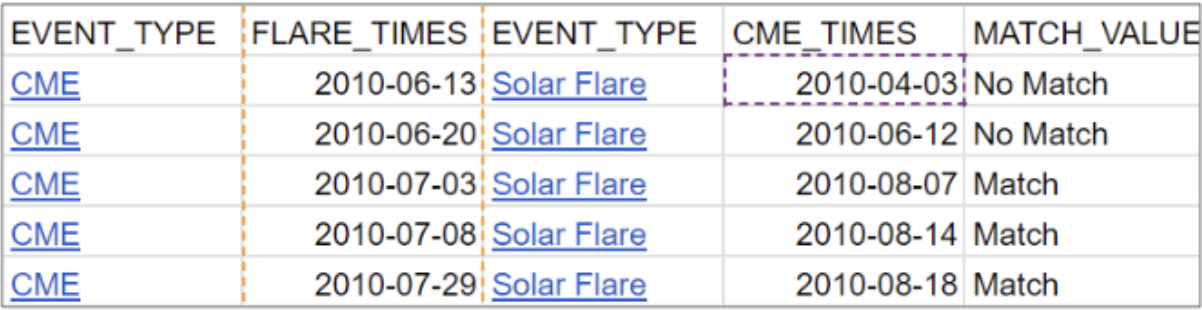}
    \caption{Visualization of matching process for CME and Solar Flare classes. Google Sheet processes would determine whether a date from CME column would match with Solar Flare date.}
    \label{fig:matching}
\end{figure}

To compile a dataset, we shift arrival dates of solar flares and CMEs back by one day, then select dates that match between solar flares and CMEs, as shown in fig. \ref{fig:matching}. We then write a Google Apps Script \cite{appscript} to synthesize a dataset by extracting the magnetogram from the date of the matched events, then upload a label of ‘1’ for the CME and solar flare, meaning that the events both occurred on the date. Then, we observe dates where there was an observed CME, but no solar flare, and extract magnetograms for these dates, but upload a label of ‘1’ for CME and ‘0’ for solar flares. 

\begin{figure}[H]
    \centering
    \includegraphics[width=\linewidth]{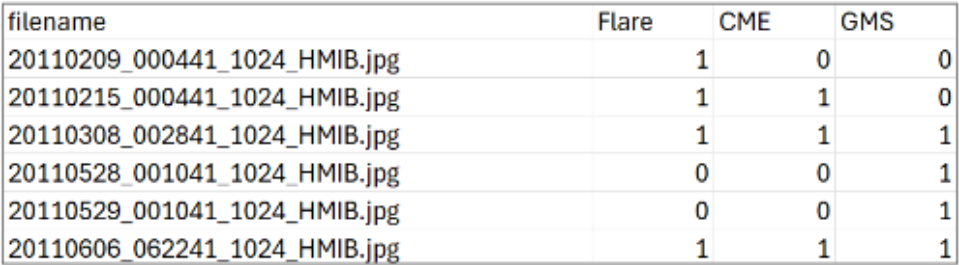}
    \caption{Dataframe example. Contains image file along with three classes, Flare, CME, and GMS, with binary labels where 1 represents event and 0 represents no event.}
    \label{fig:dataframe}
\end{figure}

Then, we observe dates where there was an observed solar flare and no observed CME, and extract magnetograms for these dates, but upload a label of ‘0’ for CME and ‘1’ for solar flares, as shown in fig. \ref{fig:dataframe}. Then, we manually input dates during which a GMS was observed, inputting ‘1’ for the GMS label. Following this, we extract magnetograms from dates (shifted back 1 day) where there was no observed solar event, uploading a ‘0’ for CME, solar flare, and GMS. Our methodology is presented in fig. \ref{fig:methods}. This original dataset is given by $D_{\text{data}}$. We apply an 80/20 train/test split to this data, then proceed to resample train data.

\subsection{Data Resampling}

Data resampling was necessary due to imbalance between classes in $D_{\text{data}}$ as a result of data collection procedures resulting in a disproportionate number of ‘1’cases for each class. 
\begin{figure}[H]
    \centering
    \includegraphics[width=0.8\linewidth]{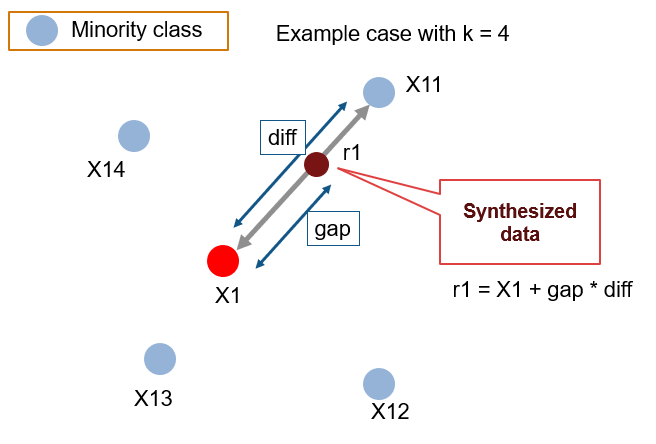}
    \caption{Visualization of the SMOTE algorithm. The SMOTE algorithm \cite{Smote} synthesizes data using real data from the minority class of the dataset. Courtesy of Analytics Vidhya.}
    \label{fig:smote}
\end{figure}
To alleviate this, we apply the Synthetic Minority Over-sampling Technique (SMOTE) algorithm, visualized in fig. \ref{fig:smote} in order to synthesize samples from the minority class, which are the ‘0’ cases for the flare and CME classes. This was done separately per class, as the GMS case had significantly more ‘0’ cases than ‘1’ cases. We further resample Ddata  using the SMOTE algorithm, as a result of an imbalance between the number of GMS observations in comparison to solar flare and CME cases.

\subsection{Model Architecture}

For this image classification problem, we utilize a custom architecture convolutional neural network (CNN) due to the process by which it is able to identify patterns in image inputs \cite{CNN}. 

A CNN is composed of multiple layers, which include the input layer, convolutional layer, pooling layer, rectified linear unit layer, and fully connected layer \cite{CNN}. 
The input layer receives the original image input, then processes and resizes the image in order to be passed to the following layers \cite{CNN}. Each pixel in the image is assigned a value from 0-255 \cite{CNN}. The convolutional layer applies a filter to the image, where a filter is a matrix of values within this range \cite{CNN}. This filter extracts features from the image, creating a ‘feature map’ of patterns in the image \cite{CNN}. Feature maps are calculated based on the formula:

\vspace{1cm} 

\textit{$G[m, n] = (f * h)[m, n] = \sum_i \sum_j h[i, j] f[m-i, n-j]$}

\vspace{1cm} 

Where \textit{f} represents a pixel in the input image, while h represents a kernel image, while \textit{m} represents the columns of the matrix, and \textit{n} represents the rows of the matrices \cite{cnnmath}. 

The pooling layer downsamples the image while preserving the ‘features’ of the image \cite{CNN}. The rectified linear unit layer, or ‘ReLU’ layer, is an activation function that implements nonlinearity into the model, which is extremely important considering the multi-label binary classification problem we aim to target \cite{CNN}. The fully connected layer converts the observed patterns to generate predictions \cite{CNN}). 

\begin{figure}[H]
    \centering
    \includegraphics[width=0.6\linewidth]{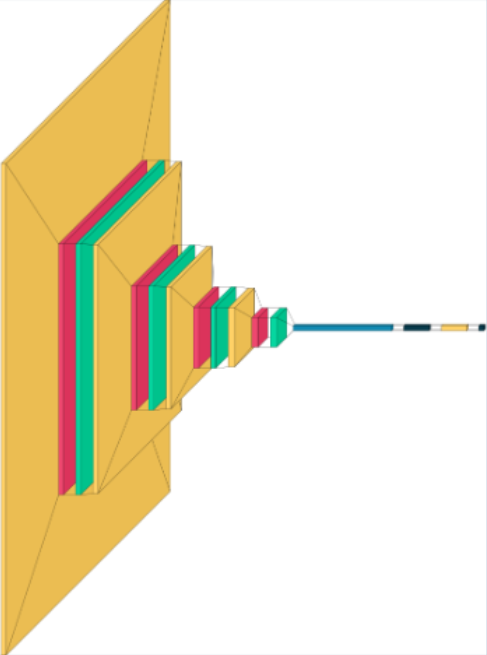}
    \caption{Visualization of Convolutional Neural Network. Contains four convolutional layers and four max pooling layers, which have provided strong results.}
    \label{fig:cnn}
\end{figure}

We propose a model using four convolutional layers and four max pooling layers using the ReLU activation function, as shown in fig. \ref{fig:cnn}.

\subsection{Model Training and Testing}

\begin{figure}[H]
    \centering
    \includegraphics[width=0.7\linewidth]{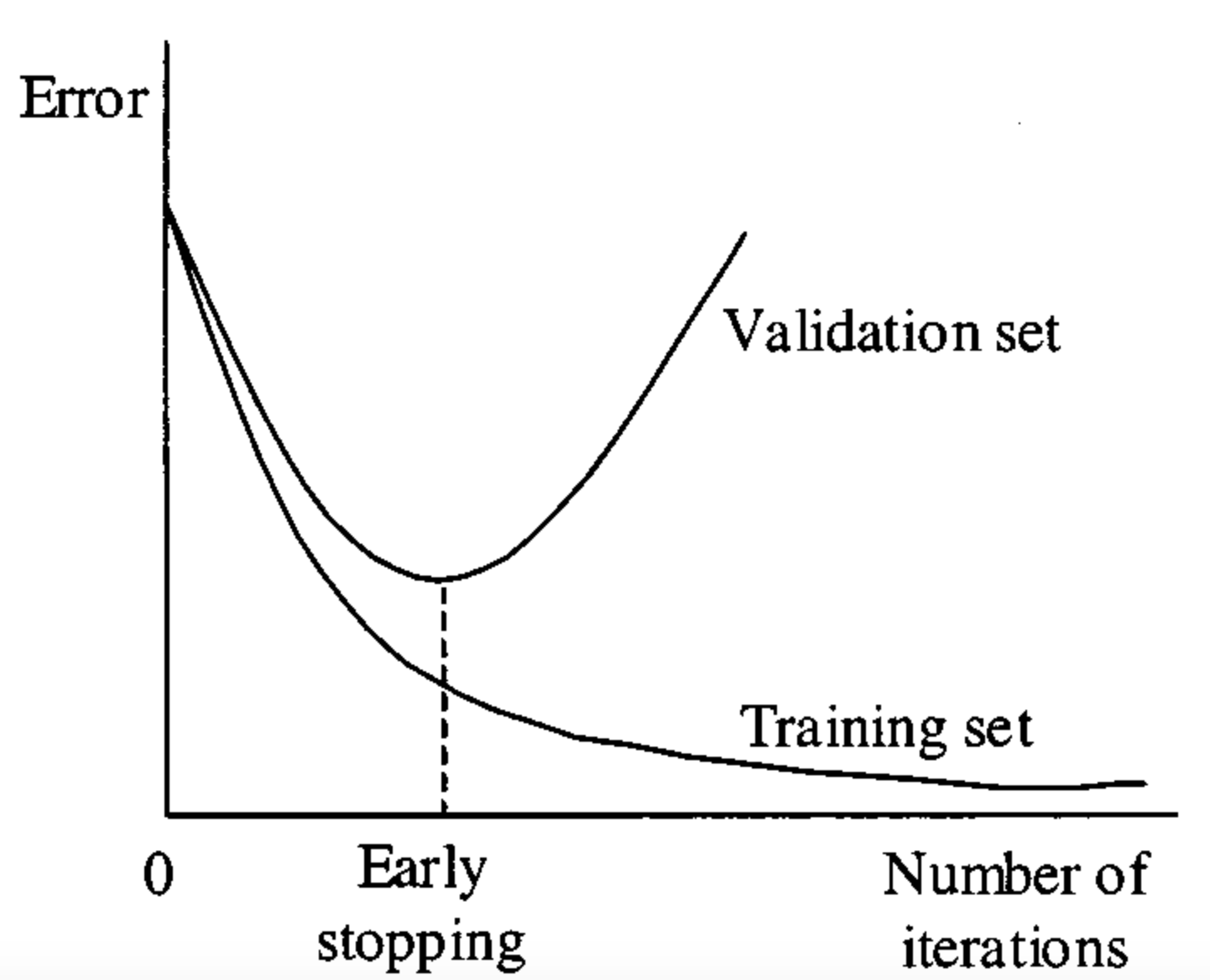}
    \caption{Visualization of early stopping mechanism. After the model converges, the early stopping mechanism halts the training process to prevent overfitting. Courtesy of Ramazan Gençay.}
    \label{fig:early}
\end{figure}

After implementing custom model architecture, the neural network was trained for 20 epochs, with an early stopping mechanism implemented to prevent overfitting as a result of the resampling that was done to the training data, as shown in fig. \ref{fig:early} \cite{Earlystop}. 

The learning rate of the network was 0.0001, using the Adam \cite{Adam} optimization framework. 


\section{Results}

\subsection{Metrics}
We evaluate all models on the metrics of precision, accuracy, recall, and F1 score. These metrics are computer as following:

\scriptsize\[
\text{accuracy} = \dfrac{\text{True Positives}}{\text{True Positives} + \text{False Positives} + \text{True Negatives}  + \text{False Negatives}},
\]
\\
\scriptsize\[
\text{precision} = \dfrac{\text{True Positives}}{\text{True Positives} + \text{False Positives}},
\]
\\
\scriptsize\[
\text{recall} = \dfrac{\text{True Positives}}{\text{True Positives} + \text{False Negatives}},
\]
\\
\scriptsize\[
\text{F1} = \dfrac{2 \cdot \text{precision} \cdot \text{recall}}{\text{precision} + \text{recall}}.
\]
\normalsize

\subsection{Evaluation}

\begin{figure}[H]
    \centering
    \includegraphics[width=0.7\linewidth]{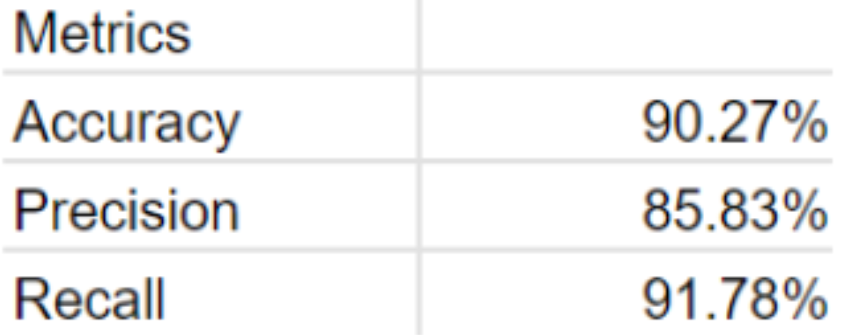}
    \caption{Table presenting accuracy, precision, and recall of model. The model performed strongly in each of these classes.}
    \label{fig:metrics}
\end{figure}

\begin{figure}[H]
    \centering
    \includegraphics[width=0.7\linewidth]{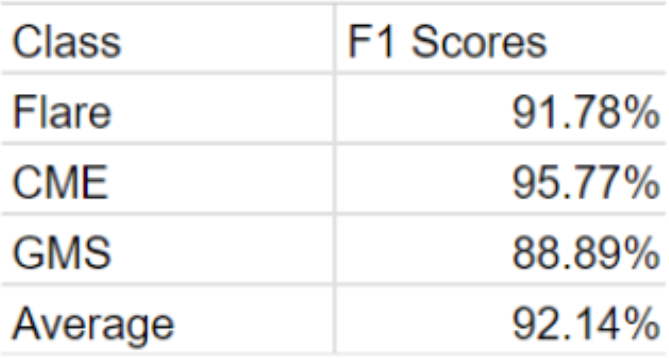}
    \caption{Table presenting F1 Scores of the model in each class. The model performed strongly in each of these categories.}
    \label{fig:f1}
\end{figure}

The model returned an accuracy of 90.27\%, precision of 85.83\%, recall of 91.78\%, and F1 score of 92.14\%, along with strong metrics for F1 scores per class, as shown in fig. \ref{fig:metrics} and fig \ref{fig:f1}.

We present confusion matrices to further visualize data, which show the proportion of true positive, false positive, true negative, and false positive predictions. We use normalized confusion matrices in order to show the true proportion of classes predicted, because there was strong imbalance between classes in the dataset. The following confusion matrices provide insight on the model's performance per class: 

\begin{figure}[H]
    \centering
    \includegraphics[width=0.8\linewidth]{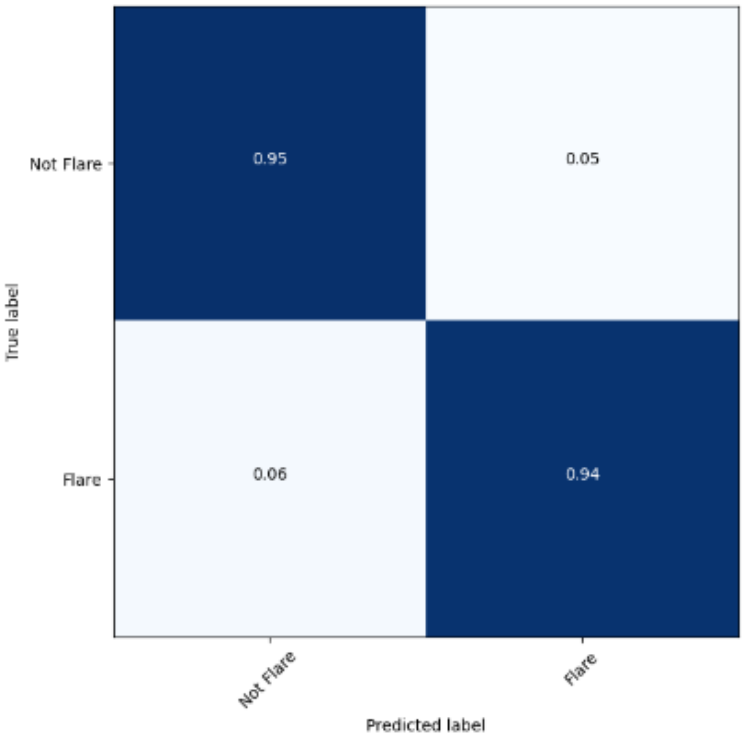}
    \caption{Confusion matrix visualizing model performance in predicting solar flare cases.}
    \label{fig:flare}
\end{figure}
\vspace{-\baselineskip}

For the solar flare class, the network performs well, as characterized by the strong downwards slope as shown in fig. \ref{fig:flare}.

\begin{figure}[H]
    \centering
    \includegraphics[width=0.8\linewidth]{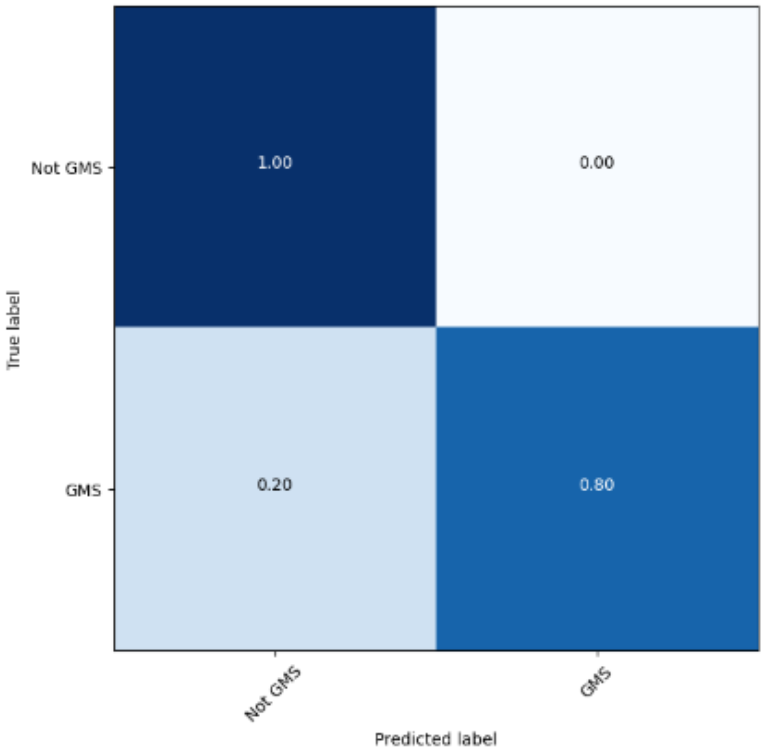}
    \caption{Confusion matrix visualizing model performance in predicting geomagnetic storm cases.}
    \label{fig:gms}
\end{figure}

For the GMS class, the network also performs relatively well, as characterized by the defined downwards slope as shown in fig. \ref{fig:gms}. However, this class had a major imbalance between event and no event data labels, and as a result, data synthesis may have resulted in overfitting to the ‘Not GMS’ class, resulting in a 1.00 accuracy in predicting the case in testing.

\begin{figure}[H]
    \centering
    \includegraphics[width=0.8\linewidth]{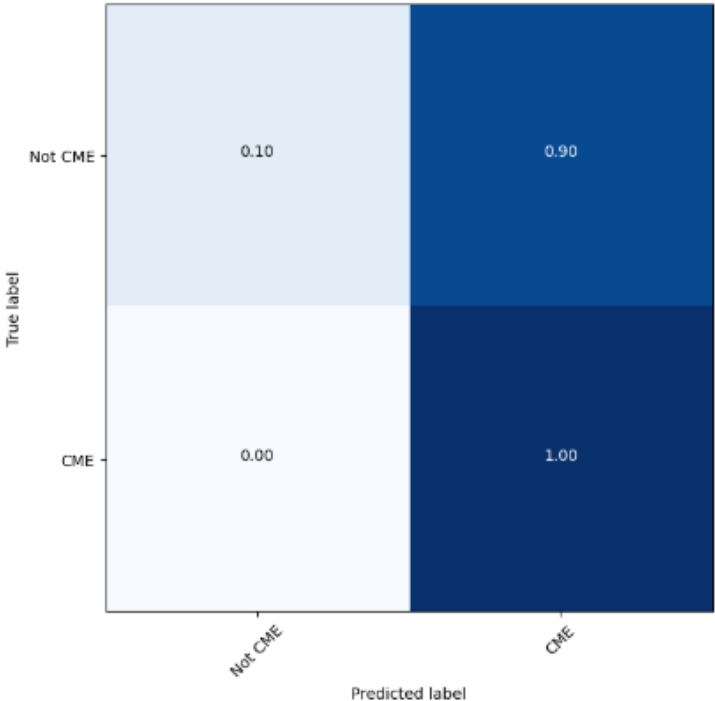}
    \caption{Confusion matrix visualizing model performance in predicting coronal mass ejection cases.}
    \label{fig:cme}
\end{figure}

For the CME class, the network performs slightly poorly, as characterized by the vertical bar observed in fig. \ref{fig:cme}. This is likely due to overfitting in the model’s training, as a result of the major disparity between event and no event cases in the training data. Though resampling was done, it relied heavily on the currently available data, which was insufficient and likely resulted in overfitting for this class. 

With strong performance in the solar flare and GMS classes, the model shows promising results in the field of space weather prediction.


\section{Discussion and Conclusions}

We propose a method of space weather prediction through the use of active region solar magnetograms. We extend the research of Domico et al., Zhang et al., and Bobra et al. to develop a system of space weather prediction which is capable of predicting multiple classes of space weather based on magnetogram inputs into a custom architecture CNN \cite{gmsforecast} \cite{flareprediction2022} \cite{CMEprediction}. Future work can involve prediction of the strengths of events in order to generate more relevant forecasts of space weather, and utilizing time-series data in order to generate a more reliable forecast based on the solar cycle. 
We find that the proposed model is capable of accurately predicting instances of solar flares and geomagnetic storms, while it requires further testing and data aggregation in order to more accurately predict instances of coronal mass ejections. As observed, the model returned an accuracy of 90.27\%, precision of 85.83\%, recall of 91.78\%. These statistics show promising results for the model, in that it is capable of making accurate and relevant predictions of when a solar event may occur, 24 hours in advance. A more uniform distribution of classes in the data would enable the model to better identify patterns in magnetograms, resulting in higher quality predictions and stronger performance. 
The proposed model and dataset aggregation method can be leveraged in further studies which explore the prediction of space weather events using artificial intelligence.


\bibliographystyle{IEEEtran}
\bibliography{refs}

\end{document}